\documentclass[12pt,a4paper]{article}
\usepackage[utf8]{inputenc}
\usepackage[T1]{fontenc}
\usepackage[margin=2.0cm]{geometry}
\usepackage[pdfstartview={FitH},bookmarks=false,linktoc=page,colorlinks=true,linkcolor=blue,citecolor=blue,urlcolor=blue]{hyperref}
\usepackage[numbered]{bookmark}
\usepackage{mathtools}
\usepackage{amssymb}
\usepackage{graphicx}
\usepackage[font=small,labelfont=bf]{caption}
\usepackage{authblk}
\usepackage{cite}
\usepackage{dsfont}
\usepackage[dvipsnames]{xcolor} 
\usepackage[titles]{tocloft}
\usepackage{float}
\usepackage{subcaption}
\usepackage{color}
\usepackage{tikz}
\usepackage{ifthen}
\usepackage[warn]{textcomp}%for the number sign
\numberwithin{equation}{section}
\allowdisplaybreaks
\usepackage{multirow}
\usepackage{cancel}

\usetikzlibrary{patterns}

\usepackage{comment}

\usepackage{caption}
\usepackage{subcaption}

\begin{document}
	%%%%%%%%%%%%%%%%%%%%%%%%%%%%%%%%%%%%%%%%%%%%
	\title{\vspace{0cm}\textbf{ Thermodynamic Stability and Fluctuations of the (2+1)-dimensional GMG Warped Black Hole }\vspace{0cm}}
	%%%%%%%%%%%%%%%%%%%%%%%%%%%%%%%%%%%%%%%%%%%%

	%\author[a]{V. Avramov}
	%\author[a,b]{H. Dimov}
	%\author[a]{M. Radomirov}
	%\author[a,c]{R. C. Rashkov}
	\author[a]{T. Vetsov\thanks{Email: vetsov@phys.uni-sofia.bg}}
	
	\affil[a]{\textit{Department of Theoretical Physics,}\authorcr\textit{Faculty of Physics,} \authorcr\textit{Sofia University ``St. Kliment Ohridski'',}\authorcr\textit{5 J. Bourchier Blvd., 1164 Sofia, Bulgaria}}
	\date{}
	\maketitle
	
	\begin{abstract}
We investigate the thermodynamic stability and the stochastic thermal fluctuations of the warped black hole solution in three-dimensional General Massive Gravity. We demonstrate that the black hole is  thermodynamically unstable and identify the nontrivial Davies phase-transition curves from the behavior of its admissible heat capacities. Going beyond the classical stability analysis, we study thermal fluctuations within a modified finite-time nonequilibrium extension of Ruppeiner's Hessian-based fluctuation theory. For a class of isentropic and isoenergetic processes, we derive exact on-shell angular momentum trajectories in the thermodynamic state space and compute the corresponding thermodynamic lengths. These quantities characterize relaxation processes between macrostates and provide an estimate of the associated relaxation times. Furthermore, we show that the thermodynamic geodesic equations do not admit constant-angular-momentum solutions, suggesting a continuous change of the black hole's angular momentum. Our results consistently reproduce the warped AdS$_3$ black hole limit of Topological Massive Gravity.
	\end{abstract}
	
	\thispagestyle{empty}
	%\pagebreak
	\tableofcontents
	
\section{Introduction}

The quest for a mathematically consistent theory of quantum gravity has extensively utilized lower-dimensional models as a theoretical laboratory. In three dimensions, General Relativity lacks local propagating degrees of freedom, rendering the bulk geometry flat or locally anti-de Sitter ($\text{AdS}_3$). To circumvent this and introduce dynamical gravitons, various massive extensions of three-dimensional gravity have been proposed. Among these, Topologically Massive Gravity (TMG) introduces a parity-odd propagating massive mode via a gravitational Chern-Simons term \cite{Deser:1981wh}, while New Massive Gravity (NMG) incorporates parity-even fourth-order curvature terms to propagate a massive spin-2 local degrees of freedom \cite{Bergshoeff:2009hq}. 

A natural unification of these two paradigms yields General Massive Gravity (GMG) \cite{Bergshoeff:2009aq}, which describes the most general combination of the Einstein-Hilbert action, a cosmological constant, the parity-odd TMG sector, and the higher-derivative NMG terms. The rich parameter space of GMG admits a variety of non-trivial vacuum configurations. Particularly compelling among these are the warped black hole solutions \cite{Tonni:2010gb}, which exhibit an asymmetric deformation of $\text{AdS}_3$ isometry groups, providing a fertile testing ground for the generalized laws of black hole mechanics and the holographic correspondence.

While the standard thermodynamic relations for these warped configurations are already well established, their thermodynamic stability using the full set of global and local conditions is nontrivial and still missing. Since higher-derivative and topologically modified theories frequently introduce non-trivial couplings that can trigger local or global instabilities, analyzing these stability profiles is crucial for understanding whether these solutions can survive thermal or quantum fluctuations or if they inevitably decay into more energetically favorable states.

In this paper, we perform a comprehensive thermodynamic study of warped black holes within the framework of three-dimensional GMG. We systematically evaluate both the global and local stability criteria \cite{Avramov:2023eif, Avramov:2024hys, Avramov:2025lmp, Avramov:2023wyi} of these configurations. Moving beyond classical steady-state thermodynamics, we utilize a modified non-equilibrium finite-time generalization \cite{Avramov:2025tlh, Avramov:2026OptimalKerr, Radomirov:2025uvj} of Ruppeiner’s thermodynamic fluctuation theory \cite{Ruppeiner1983, Ruppeiner1995} to investigate the stochastic thermal paths connecting various macrostates of the system. By analyzing the corresponding geodesic equations on the thermodynamic state space, we explore the dynamic time evolution constrained by the underlying geometry of the state space.

The  paper is organized as follows. In Section \ref{secWBHGMG}, we review the warped black hole metric in GMG and establish its basic geometric properties. In Section \ref{secTDSA}, we construct the mass-energy representation of the black hole's thermodynamics and apply the Sylvester criterion to determine global stability. We further analyze the properties of the admissible local heat capacities to obtain the corresponding Davies phase transition curves. In Section \ref{secTDLFluc} we study the stochastic thermal fluctuations along paths with minimal dissipation of the warped GMG black hole and solves the thermodynamic geodesic equations in the case of issentropic and isoenergetic change of the angular momentum. We also estimate the corresponding thermodynamic lengths and estimate the relaxation time of the processes. Finally, we provide concluding remarks and avenues for future research in Section \ref{secConcl}.
	
\section{The warped black hole in 3D general massive gravity}\label{secWBHGMG}

We review the geometric and thermodynamic properties of warped black hole solutions in three-dimensional General Massive Gravity (GMG). In particular, we determine the constraints on both the spacetime parameters and the coupling constants of the theory that are necessary to ensure regularity, causal consistency, and the absence of pathological features. 

\subsection{The warped GMG black hole metric}

The action defining General Massive Gravity (GMG) in three dimensions is given by\footnote{We adopt the mostly plus signature convention $(-, +, +)$ for the metric tensor.}:
\begin{equation}\label{eqActionGMG}
	A = \frac{1}{16 \pi G}\int d^3 x \sqrt{-g}\left(\sigma R - 2 \lambda m^2 + \frac{1}{\mu} \mathcal{L}_{\text{CS}} + \frac{1}{m^2} \mathcal{L}_{\text{NMG}}\right),
\end{equation}
where $\mathcal{L}_{\text{CS}}$ and $\mathcal{L}_{\text{NMG}}$ denote the Chern-Simons (CS) and New Massive Gravity (NMG) Lagrangian densities, respectively:
\begin{equation}
	\mathcal{L}_{\text{CS}} = \frac{1}{2}\varepsilon^{\mu\nu\rho} \left(\Gamma^\alpha_{\mu\beta}\partial_\nu\Gamma^\beta_{\rho\alpha} + \frac{2}{3}\Gamma^{\alpha}_{\mu\beta}\Gamma^{\beta}_{\nu\gamma}\Gamma^\gamma_{\rho\alpha}\right),\quad 
	\mathcal{L}_{\text{NMG}} = R_{\mu\nu}R^{\mu\nu} - \frac{3}{8} R^2.
\end{equation}
The discrete parameter $\sigma = \pm 1$ in the action \eqref{eqActionGMG} preserves the freedom to specify the sign of the Einstein-Hilbert term\footnote{A discussion regarding the physical sign  of $\sigma$ can be found in \cite{Bergshoeff:2009aq}.}. The cosmological constant $\lambda$ is dimensionless, whereas the NMG and CS coupling parameters, $m$ and $\mu$, possess dimensions of mass. 

The resulting field equations of motion admit a warped black hole solution characterized by the line element \cite{Tonni:2010gb}:
\begin{equation}
	\frac{ds^2}{\ell^2} = -N(r) dt^2 + \frac{dr^2}{4 R(r) N(r)} + R(r) \big(d\theta + D(r) dt\big)^2,
\end{equation}
where the  metric functions are defined as:
\begin{align}
	&R(r) = \frac{\zeta^2}{4} r \Big((1-\eta^2) r + \eta^2 (r_+ + r_-) + 2 \eta \sqrt{r_+ r_-}\Big),
	\\[5pt]
	&N(r) = \zeta^2 \eta^2 \frac{(r-r_+)(r-r_-)}{4 R(r)},\quad 
	D(r) = \frac{r + \eta\sqrt{r_+ r_-}}{2 R(r)}\,|\zeta|.
\end{align}
The range of the spacetime coordinates is: $-\infty < t < +\infty$, $0 \leq r < +\infty$, and $\theta \sim \theta + 2 \pi$. The geometry has two horizons located at $r_+$ and $r_-$, with the vacuum state recovered in the limit $r_+ = r_- = 0$. 

This metric solves the field equations provided that the metric parameters $(\ell,\zeta,\eta)$ are related to the coupling constants of the action $(\sigma, \mu, m, \lambda)$ via the constraints \cite{Tonni:2010gb}:
\begin{align}
	&\eta^2 = \frac{21}{4} - \frac{3 m^2 \ell}{\mu\zeta} + \frac{2 \sigma m^2\ell^2}{\zeta^2},
	\\[5pt]
	&\frac{21}{m^4 \ell^4}\zeta^4 - \frac{32}{m^2\ell^3\mu}\zeta^3 + \frac{12}{\ell^2}\left(\frac{4\sigma}{m^2} + \frac{1}{\mu^2}\right)\zeta^2 - \frac{32 \sigma}{\mu \ell}\zeta + 16(1+\lambda) = 0.
\end{align}
These relations imply that two of the four coupling constants $(\sigma, \mu, m, \lambda)$ can be explicitly integrated out in favor of the geometric parameters $(\ell, \eta, \zeta)$. In the subsequent analysis, we eliminate the pair $(\lambda, \mu)$, thereby leaving $(m, \sigma)$ as independent free parameters:
\begin{align}
	&\mu = \frac{12 \zeta m^2 \ell}{\zeta^2 \left(21 - 4 \eta^2\right) + 8 m^2 \sigma \ell^2},
	\\[5pt] 
	&\lambda = \frac{\left(\zeta^2 \left(3 - 4 \eta^2\right) + 8 m^2 \sigma \ell^2\right) \left(\zeta^2 \left(4 \eta^2 - 7\right) + 24 m^2 \sigma \ell^2\right)}{192 m^4 \ell^4} - 1.
\end{align}
Due to the invariance of the system under the inversion $\zeta \to -\zeta$, we restrict our attention to the $\zeta > 0$ domain without loss of generality. 

To verify that the spacetime is devoid of curvature singularities, we evaluate the main algebraic and differential invariants of the metric:
\begin{align}
	&R = \frac{\zeta^2 \left(1 - 4 \eta^2\right)}{2 \ell^2},
	\\[5pt]
	&R^{ab}{}_{cd} R^{cd}{}_{ab} = \frac{\zeta^4 \left(16 \eta^4 - 24 \eta^2 + 11\right)}{4 \ell^4},
	\\[5pt]
	&R^{ab}{}_{cd} R^{ef}{}_{ab} R^{cd}{}_{ef} = \frac{\zeta^6 \left(25 - 64 \eta^6 + 144 \eta^4 - 108 \eta^2\right)}{8 \ell^6},
	\\[5pt]
	&\nabla_e R_{abcd} \nabla^e R^{abcd} = -\frac{4 \zeta^6 \left(\eta^2 - 1\right)^2}{\ell^6}.
\end{align}
Furthermore, enforcing the global exclusion of closed timelike curves strictly restricts the parameter space to $0 < \eta^2 < 1$ \cite{Tonni:2010gb}. Consequently, the physically viable warped black hole is uniquely defined within the domain $\zeta > 0$ and $0 < \eta^2 < 1$, where all associated curvature invariants remain manifestly regular.

\subsection{Thermodynamics}

Under the regularity conditions $\zeta > 0$ and $0 < \eta^2 < 1$, the thermodynamic profile of the GMG warped black hole is described by the expressions \cite{Tonni:2010gb}:
\begin{align}
	&M = \frac{\zeta^2 \eta^2}{48 G m^2}\left(6\zeta^2 - m^2\right)\left(r_- + r_+ + 2\eta \sqrt{r_- r_+}\right),
	\\[5pt]
	&S = \frac{\pi \zeta \ell}{12 G m^2}\left(\left(6\zeta^2 + \left(\eta^2 - 1\right)m^2\right)r_+ - \eta^2 m^2 r_- + \eta \left(6\zeta^2 - m^2\right)\sqrt{r_- r_+}\right),
	\\[5pt]
	&J = \frac{\zeta^3 \eta^2 \ell}{16 G}\left[\left(\frac{\zeta^2}{m^2} - \frac{1}{6}\right)\left(\left(\eta^2 + 1\right)r_- r_+ + \eta \left(r_- + r_+\right)\sqrt{r_- r_+}\right) - \frac{\eta^2}{12}\left(r_- - r_+\right)^2\right],
	\\[5pt]
	&T = \frac{\zeta \eta^2\left(r_+ - r_-\right)}{4\pi \ell \left(r_+ + \eta \sqrt{r_- r_+}\right)},\quad 
	\Omega = \frac{2}{\zeta \ell \left(r_+ + \eta \sqrt{r_- r_+}\right)},
\end{align}
where $M$, $S$, $J$, $T$, and $\Omega$ represent the mass, entropy, angular momentum, Hawking temperature, and angular velocity, respectively. The thermodynamic variables satisfy both the first law of thermodynamics  and the Smarr relation:
\begin{equation}
	dM = T dS + \Omega dJ,\quad 	M = TS + 2 \Omega J.
\end{equation}

To establish a direct comparison with the Topologically Massive Gravity (TMG) results from \cite{Anninos:2008fx, Dimov:2019fxp}, we restrict our analysis to the $-1<\eta < 0$ sector. The exact TMG thermodynamic limits are recovered by the following parametric scaling:
\begin{equation}\label{eqTMGlimit}
	\zeta \to 2\nu ,\quad 
	\eta \to -\sqrt{\frac{\nu^2 + 3}{4\nu^2}} ,\quad 
	m^2 \to 8 \nu^2.
\end{equation}

Consequently, within the sector $-1 < \eta < 0$ (where we parameterize $\eta = -|\eta|$), the thermodynamic variables scale as:
\begin{align}
	&M = \frac{\zeta^2 \eta^2 \kappa}{48 G m^2}\left(r_- + r_+ - 2|\eta|\sqrt{r_- r_+}\right),
	\\[5pt]
	&S = \frac{\pi \zeta \ell}{12 G m^2}\left(\left(\kappa + \eta^2 m^2\right)r_+ - \eta^2 m^2 r_- - |\eta|\kappa \sqrt{r_- r_+}\right),
	\\[5pt]
	&J = \frac{\zeta^3 \eta^2 \ell}{192 G m^2}\left(2\kappa \left(\left(\eta^2 + 1\right)r_- r_+ - |\eta|\left(r_- + r_+\right)\sqrt{r_- r_+}\right) - \eta^2 m^2\left(r_- - r_+\right)^2\right),
	\\[5pt]
	&T = \frac{\zeta \eta^2\left(r_+ - r_-\right)}{4\pi \ell \left(r_+ - |\eta|\sqrt{r_- r_+}\right)},\quad 
	\Omega = \frac{2}{\zeta \ell \left(r_+ - |\eta| \sqrt{r_- r_+}\right)},
\end{align}
where we have introduced an auxiliary positive parameter $\kappa \equiv 6\zeta^2 - m^2 > 0$. 

The valid range of these coupling parameters is tightly constrained by thermodynamic consistency conditions. Requiring positivity of the total mass ($M > 0$) imposes the upper bound $m < \sqrt{6} \zeta$. Furthermore, ensuring a positive-definite entropy ($S > 0$) enforces a geometric restriction on the ratio of the horizon radii:
\begin{equation}
	\frac{r_+}{r_-} > \frac{\eta^2\left(36\zeta^4 + (2\eta^2 - 1)m^4 + \kappa \sqrt{36\zeta^4 + 12\zeta^2 m^2 + \left(4\eta^2 - 3\right)m^4}\right)}{2\left(\kappa + \eta^2 m^2\right)^2},
\end{equation}
which is identically satisfied throughout the selected parameter space. Finally, the extremal case $r_- = r_+$ (implying $T \to 0$) is excluded  due to violation of the third law of thermodynamics\footnote{Extremal configurations remain under active investigation within string theory and alternative approaches to quantum gravity, indicating that they should be discarded primarily at the classical level. For a detailed discussion on the mechanisms underlying the violation of the third law in black hole systems, see \cite{Davies:1978zz}.}.

\section{Classical thermodynamic stability analysis}\label{secTDSA}

We demonstrate that the three-dimensional GMG warped black hole is globally unstable from a classical thermodynamic perspective. Consequently, we derive the explicit expressions for the admissible heat capacities of the system and obtain the Davies phase transition curves.

\subsection{Mass-energy representation}

To implement the framework of classical global thermodynamic stability, the mass-energy potential $M$ must be expressed exclusively in terms of its natural extensive variables, namely the entropy $S$ and the angular momentum $J$. To decouple the thermodynamic degrees of freedom, we parameterize the horizon radii via the scaling relation $r_+ = b^2 r_-$, where the dimensionless parameter satisfies $b > 1$. By substituting this ansatz into the black hole's thermodynamics, one can explicitly solve for $r_+$ and $r_-$ while adhering to the geometric constraints established in the preceding section. Following a straightforward yet tedious algebraic computation, the mass-energy representation yields:
\begin{align}
	M(S,J) &= \frac{\zeta \kappa}{12\pi G\ell m^2}\left( \sqrt{\frac{3G\left(\kappa + 2\eta^2 m^2\right)\left(3\zeta G S^2 + 4\pi^2 J\ell\right)}{\zeta \kappa}} - 3GS \right), \label{eqMassEnsemble}
	\\[5pt]
	T(S,J) &= \left.\frac{\partial M}{\partial S}\right|_{J} = \frac{\zeta}{4\pi \ell m^2}\left( \frac{S \sqrt{3\zeta G\kappa \left(\kappa + 2\eta^2 m^2\right)}}{\sqrt{3\zeta G S^2 + 4\pi^2 J\ell}} - \kappa \right),
	\\[5pt]
	\Omega(S,J) &= \left.\frac{\partial M}{\partial J}\right|_{S} = \frac{\pi \sqrt{\zeta \kappa \left(\kappa + 2\eta^2 m^2\right)}}{2m^2\sqrt{3G\left(3\zeta G S^2 + 4\pi^2 J\ell\right)}}.
\end{align}

Imposing the physical requirements that both the total mass-energy and the Hawking temperature remain strictly positive ($M > 0$ and $T > 0$), we obtain a well-defined bound on the ratio of the angular momentum and the square of the entropy:
\begin{equation}\label{eqBHExists}
	-\frac{3\zeta G\eta^2 m^2}{2\pi^2\ell \left(\kappa + 2\eta^2 m^2\right)} < \frac{J}{S^2} < \frac{3\zeta G\eta^2 m^2}{2\pi^2\ell \kappa}.
\end{equation}
The saturation of these bounds defines the extremal limits of the angular momentum:
\begin{equation}\label{eqExtrJGMG}
	J_{\text{extr}}^{(-)} = -\frac{3\zeta G\eta^2 m^2}{2\pi^2\ell \left(\kappa + 2\eta^2 m^2\right)} S^2,\quad	J_{\text{extr}}^{(+)} = \frac{3\zeta G\eta^2 m^2}{2\pi^2\ell \kappa} S^2.
\end{equation}

Notably, by taking the Topologically Massive Gravity (TMG) limit defined in Eq.~\eqref{eqTMGlimit}, we smoothly recover the established TMG constraint obtained in \cite{Avramov:2025zxc}:
\begin{equation}
	-\frac{3G\nu \left(\nu^2 + 3\right)}{2\pi^2\ell \left(5\nu^2 + 3\right)} < \frac{J}{S^2} < \frac{3G\left(\nu^2 + 3\right)}{8\pi^2\ell \nu}.
\end{equation}

\subsection{Global thermodynamic instability}

The global thermodynamic stability or instability of the black hole  is determined by the convexity of its mass-energy potential. Mathematically, this property is encoded in the Hessian matrix of the mass with respect to its extensive variables. According to the Sylvester criterion, a state is thermodynamically stable if and only if the Hessian matrix $\mathcal{H}_M$ is positive-definite, which requires all of its principal minors to be strictly positive \cite{Avramov:2023eif, Avramov:2024hys, Avramov:2025lmp, Avramov:2023wyi}. For a two-dimensional state space this yields:
\begin{equation}
	\left.\frac{\partial^2 M}{\partial S^2}\right|_J > 0,\quad 
	\left.\frac{\partial^2 M}{\partial J^2}\right|_S > 0,\quad 
	\det \mathcal{H}_M = \left.\frac{\partial^2 M}{\partial S^2}\right|_J \left.\frac{\partial^2 M}{\partial J^2}\right|_S - \left(\frac{\partial^2 M}{\partial S\partial J}\right)^{2} > 0.
\end{equation}

For the GMG warped black hole the explicit evaluation of the mass-energy Hessian yields:
\begin{align}
	\mathcal{H}_M &= 
	\begin{pmatrix}
		\left.\frac{\partial^2 M}{\partial S^2}\right|_J & \frac{\partial^2 M}{\partial S\partial J} \\[12pt]
		\frac{\partial^2 M}{\partial S\partial J} & \left.\frac{\partial^2 M}{\partial J^2}\right|_S
	\end{pmatrix} 
	= \begin{pmatrix}
		\frac{\sqrt{3} \pi J \sqrt{\zeta^3 G \kappa \left(2 \eta^2 m^2 + \kappa \right)}}{m^2 \left(3 \zeta G S^2 + 4 \pi^2 J \ell\right)^{3/2}} & -\frac{\sqrt{3} \pi S \sqrt{\zeta^3 G \kappa \left(2 \eta^2 m^2 +\kappa \right)}}{2 m^2 \left(3 \zeta G S^2 + 4 \pi^2 J \ell\right)^{3/2}} \\[12pt]
		-\frac{\sqrt{3} \pi S \sqrt{\zeta^3 G \kappa \left(2 \eta^2 m^2 + \kappa \right)}}{2 m^2 \left(3 \zeta G S^2 +4 \pi^2 J \ell\right)^{3/2}} & -\frac{\pi^3 \ell \sqrt{\zeta \kappa \left(2 \eta^2 m^2 + \kappa \right)}}{\sqrt{3 G} m^2 \left(3 \zeta G S^2 +4 \pi^2 J \ell\right)^{3/2}}
	\end{pmatrix}.
\end{align}
Direct inspection reveals that the Hessian matrix fails to be positive-definite. Specifically, its determinant is strictly negative throughout the entire physically allowed domain:
\begin{equation}
	\det \mathcal{H}_M = -\frac{\pi^2 \zeta^2 \kappa \left(\kappa + 2\eta^2 m^2\right)}{4m^4\left(3\zeta G S^2 + 4\pi^2 J\ell\right)^2} < 0.
\end{equation}
The violation of the Sylvester criterion implies that the GMG warped black hole cannot attain global thermodynamic equilibrium. Instead, the geometry represents an intrinsically non-equilibrium  system prone to continuous energy radiation.

In the TMG limit~\eqref{eqTMGlimit}, the Hessian determinant reduces to:
\begin{equation}
	\det \mathcal{H}_M^{\text{TMG}} = -\frac{\pi^2 \left(5 \nu^2 + 3\right)}{4 \left(3 G \nu S^2 + 2 \pi^2 J \ell\right)^2} < 0,
\end{equation}
which precisely reproduces the global thermodynamic instability profile established for warped $\text{AdS}_3$ black holes in Topologically Massive Gravity \cite{Dimov:2019fxp}.

\subsection{Local thermodynamic stability and phase transitions}

The global thermodynamic instability of the system implies that its local response functions --- specifically its heat capacities --- possess negative values in the admissible parameter regimes. To characterize the local stability properties and locate potential Davies phase transitions, we evaluate the heat capacities under different thermodynamic constraints. Utilizing the standard Nambu bracket formalism \cite{Mansoori:2014oia, Avramov2023Thermodynamic}, the three independent heat capacities for the GMG  black hole yield:
\begin{align}
	C_{\Omega} &= T\left.\frac{\partial S}{\partial T}\right|_{\Omega} = T \frac{\{S, \Omega\}_{S,J}}{\{T, \Omega\}_{S,J}} = S - \frac{\sqrt{\kappa \left(3 \zeta G S^2 + 4 \pi^2 J \ell \right)}}{\sqrt{3\zeta G \left(\kappa + 2 \eta^2 m^2\right)}},
	\\[5pt]
	C_J &= T\left.\frac{\partial S}{\partial T}\right|_J = T \frac{\{S, J\}_{S,J}}{\{T, J\}_{S,J}} = \frac{3 \zeta G S^2 + 4 \pi^2 J \ell}{4 \pi^2 J \ell}\left(S - \sqrt{\frac{\kappa \left(3 \zeta G S^2 + 4 \pi^2 J \ell \right)}{3\zeta G \left(\kappa + 2 \eta^2 m^2\right)}}\right),
	\\[5pt]
	C_M &= T\left.\frac{\partial S}{\partial T}\right|_M \!\!= T \frac{\{S, M\}_{S,J}}{\{T, M\}_{S,J}} = S - \frac{4 \pi^2 J \ell}{\sqrt{\dfrac{3 \zeta G}{\kappa}\left(\kappa + 2 \eta^2 m^2\right) \left(3 \zeta G S^2 + 4 \pi^2 J \ell \right)} - 3 \zeta G S}.
\end{align}

An analysis of the response function $C_\Omega$ reveals that it vanishes ($C_\Omega = 0$) along the positive extremal curve $J\to J_{\text{extr}}^{(+)}$ where $T \to 0$, and diverges ($C_\Omega \to \infty$) in the limit $\zeta \to 0$. Consequently, the Davies phase transition curves for $C_\Omega$ occur for $\zeta \to 0$, $\zeta \to \infty$, and $J \to J_{\text{extr}}^{(+)}$. Within the physically permitted existence domain defined in Eq.~\eqref{eqBHExists}, $C_\Omega$ remains positive-definite.

A similar look at $C_J$ reveals that it vanishes along the extremal boundary $J\to J_{\text{extr}}^{(+)}$ and exhibits critical divergences when $\zeta \to 0$ and $\zeta \to \infty$.Furthermore,  $C_J$ undergo a sign inversion across the static black hole configuration ($J=0$):
\begin{equation}
	C_J = \begin{cases} 
		> 0, & \text{for } 0 < J < J_{\text{extr}}^{(+)}, \\[5pt]
		\to \pm\infty, & \text{for } J\to 0^{\pm},\\[5pt]
		< 0, & \text{for } J_{\text{extr}}^{(-)} < J < 0,
	\end{cases}
\end{equation}
identifying $J=0$ as the fourth Davies point. This suggests that there are two different sectors --- one for $0<J<J_{\text{extr}}^{(+)}$ and one for $J_{\text{extr}}^{(-)}<J<0$.

Finally, the heat capacity $C_M$ remains strictly positive for the physically allowed parameter space. It becomes zero at the extremal limit $J \to J_{\text{extr}}^{(+)}$, and diverges when $\zeta \to 0$.

To summarize, the four Davies phase transition curves for the GMG warped black hole are $\zeta\to 0,\infty$, $J\to 0$ and $J \to J_{\text{extr}}^{(+)}$.

\section{Thermodynamic length and fluctuations}\label{secTDLFluc}

Moving beyond classical stability analysis, we can consider the stochastic nature of transitions between thermodynamic states. As established in the previous section, the GMG warped black hole is intrinsically unstable, placing it inherently in a non-equilibrium state. Consequently, the system is expected to undergo spontaneous transitions toward more probable macrostates. We aim to quantify the probability of such fluctuations within a finite-time geometric framework, developed in \cite{Avramov:2025tlh}. To this end, we employ Ruppeiner's fluctuation theory \cite{Ruppeiner1983, Ruppeiner1995}.

\subsection{General setup}

In Ruppeiner's fluctuation theory, the probability of a fluctuation is related to the thermodynamic length $\mathcal{L}$ of the geodesic connecting two equilibrium states on a thermodynamic manifold. In the mass-energy representation, the squared length $\mathcal{L}^2$ carries units of energy, consistent with its interpretation as the minimum work required to drive the system between states. To account for the non-equilibrium nature of the Hessian, we follow \cite{Avramov:2025tlh} by introducing a scale factor $\epsilon \in \mathbb{R}$ into the thermodynamic metric. The underlying geometry is thus defined by the Weinhold metric, which is given by the rescaled Hessian of the mass potential: $g^{(W)} = \epsilon \mathcal{H}_M$.

For a general thermodynamic potential $\Phi(\Phi^1, \dots, \Phi^n)$ as a function of its natural extensive parameters $\Phi^a$, $a=1,...,n$, the thermodynamic length $\mathcal{L}$ along a path $\gamma$ can be defined in two equivalent ways. The first is the unparameterized (direct) line element:
\begin{equation}
	\mathcal{L}_{\text{direct}} = \int_\gamma \sqrt{g_{ab} d\Phi^a d\Phi^b}.
\end{equation}
The second definition utilizes an affine parameter $t$ to describe the evolution of the coordinates $\Phi^a(t)$:
\begin{equation}\label{eqGeodAction2}
	\mathcal{L}_{\text{on-shell}} = \int_{0}^{\tau} \sqrt{g_{ab}\big(\vec \Phi(t)\big)\dot \Phi^a(t) \dot \Phi^b(t)}\, dt.
\end{equation}

Extremizing this functional leads to the thermodynamic geodesic equations:
\begin{equation}\label{eqGeodesicGeneric}
	\ddot \Phi^c(t) + \Gamma^c_{ab} \dot \Phi^a(t) \dot \Phi^b(t) = 0,
\end{equation}
where the Christoffel symbols $\Gamma^{c}_{ab}$ are derived from the thermodynamic metric $g_{ab}$ in the standard manner:
\begin{equation}\label{eqChristofel}
	\Gamma^{c}_{ab} = \frac{1}{2}g^{cd}(\partial_a g_{db} + \partial_b g_{da} - \partial_d g_{ab}).
\end{equation}
The resulting geodesic profiles $\Phi^a(t)$ define the optimal transformation paths between macrostates. Physically, a larger on-shell thermodynamic length corresponds to a lower probability of a spontaneous fluctuation along that specific trajectory. This is also related to the standard interpretation of the quantity $\mathcal{L}^2$, which has units of energy (mass) and corresponds to the minimal energy required to drive the system from one thermodynamic state to another, i.e. the system will evolve along the path with the minimal resistance. 

\subsection{Geodesic equations in $(S,J)$ space and fluctuations}

For the GMG warped black hole, the state space is spanned by the entropy $S$ and angular momentum $J$. Under the metric $g^{(W)} = \epsilon \mathcal{H}_M$, the geodesic equations for these parameters are:
\begin{align}\label{eqSener}
	&\ddot S + \Gamma^S_{SS} \dot S^2 + 2\Gamma^S_{S J} \dot S \dot J + \Gamma^S_{JJ} \dot J^2 = 0, \\[5pt]\label{eqJener}
	&\ddot J + \Gamma^J_{JJ} \dot J^2 + 2\Gamma^J_{S J} \dot S \dot J + \Gamma^J_{S S} \dot S^2 = 0,
\end{align}
with the Christoffel symbols given by:
\begin{align}
	&\Gamma^{S}_{SS} = -\frac{3 \zeta G S}{3 \zeta G S^2 + 4 \pi^2 \ell J}, 
	\quad \Gamma^{S}_{SJ} = -\frac{\pi^2 \ell}{3 \zeta G S^2 + 4 \pi^2 \ell J}, \quad \Gamma^{S}_{JJ} = 0, \\[5pt]
	&\Gamma^{J}_{SS} = \frac{3 \zeta G J}{3 \zeta G S^2 + 4 \pi^2 \ell J}, \quad \Gamma^{J}_{SJ} = -\frac{3 \zeta G S}{3 \zeta G S^2 + 4 \pi^2 \ell J}, \quad \Gamma^{J}_{JJ} = -\frac{3 \pi^2 \ell}{3 \zeta G S^2 + 4 \pi^2 \ell J}.
\end{align}

It is important to note that the Weinhold thermodynamic curvature is regular everywhere: 
\begin{equation}
R^{W}=	\frac{4 \sqrt{3} \pi  m^2  G \ell }{\epsilon  \sqrt{\zeta  G \kappa  \left(\kappa +2 \eta ^2 m^2\right) \left(3 \zeta  G S^2+4 \pi ^2 J \ell \right)}},
\end{equation}
thus it does not further restrict the ($S,J$) parameter space. Note also that its sign is governed by the sign of the metric scale parameter $\epsilon$. For $\epsilon>0$ one has elliptic information space, while for $\epsilon<0$ --- a hyperbolic information geometry occurs. 

Although the coupled geodesic equations above generally require numerical treatment, analytical solutions exist for specific cases. For instance, for an isentropic fluctuation ($S = S_0 = \text{const}$), Eq.~\eqref{eqSener} is satisfied identically, and Eq.~\eqref{eqJener} reduces to:
\begin{equation}
	\ddot J(t) - \frac{3 \pi^2 \ell}{3 \zeta G S_0^2 + 4 \pi^2 \ell J(t)} \dot J(t)^2 = 0, \quad J(0) = J_0, \quad \dot J(0) = \dot J_0,
\end{equation}
where $J_0$ and $\dot J_0$ denote the initial angular momentum and its rate of change. The solution is a fourth degree polynomial in  time:
\begin{equation}\label{eqJtProfileEnergyRep}
	J_{ij}(t) = \pm |J_0|\pm |\dot J_0| t+\frac{3 \pi ^2 \dot J_0^2  \ell }{6 \zeta  G S_0^2\pm8 \pi ^2 |J_0| \ell }t^2\pm\frac{\pi ^4 |\dot J_0|^3  \ell ^2}{\left(3 \zeta  G S_0^2\pm4 \pi ^2 |J_0| \ell \right)^2}t^3+\frac{\pi ^6 \dot J_0^4  \ell ^3}{4 \left(3 \zeta  G S_0^2\pm4 \pi ^2 |J_0| \ell \right)^3}t^4,
\end{equation}
where, depending on the signs of the initial angular momentum and its rate, one has four solutions:
\begin{equation}
	J_{ij} = \begin{cases} 
	J_{++}, & \text{for } J_0=+|J_0|>0, \text{ and } \dot J_0=+|\dot J_0|>0, \\[5pt]
		J_{+-}, & \text{for } J_0=+|J_0|>0, \text{ and } \dot J_0=-|\dot J_0|<0, \\[5pt]
			J_{-+}, & \text{for } J_0=-|J_0|<0, \text{ and } \dot J_0=+|\dot J_0|>0, \\[5pt]
				J_{--}, & \text{for } J_0=-|J_0|<0, \text{ and } \dot J_0=-|\dot J_0|<0 .
	\end{cases}
\end{equation}

Substituting the on-shell trajectories into the thermodynamic length functional, we find that only $J_{++}(t)$ and $J_{+-}(t)$, corresponding to the region $0<J<J_{\text{extr}}^{(+)}$, give rise to non-negative thermodynamic lengths\footnote{This observation further corroborates the existence of two thermodynamically distinct sectors of the model, namely $0<J<J_{\text{extr}}^{(+)}$ and $J_{\text{extr}}^{(-)}<J<0$.}. Their associated thermodynamic lengths will be denoted by $\mathcal{L}{\text{on-shell}}^{(+)}(\tau)$ and $\mathcal{L}{\text{on-shell}}^{(-)}(\tau)$, respectively:
\begin{align}
	\mathcal{L}_{\text{on-shell}}^{(+)}(\tau) &= 2 \sqrt{-\epsilon} \frac{\sqrt[4]{\zeta \kappa (\kappa + 2 \eta^2 m^2) (3 \zeta G S_0^2 + 4 \pi^2 \ell |J_0| )}}{m \sqrt[4]{3 G} \sqrt{\ell |\dot J_0| \big(3 \pi \zeta G S_0^2 + \pi^3 \ell\big(4 |J_0|+|\dot J_0| \tau\big)\big)}} \nonumber \\
	&\times \left(\sqrt{3 \zeta G S_0^2 + \pi^2 \ell \big(4 |J_0|+|\dot J_0| \tau\big)} - \sqrt{3 \zeta G S_0^2 + 4 \pi^2 \ell |J_0|}\right),\quad \epsilon<0,
\end{align}
valid for $\epsilon < 0$ and:
\begin{align}
	\mathcal{L}_{\text{on-shell}}^{(-)}(\tau) &= 2 \sqrt{\epsilon} \frac{\sqrt[4]{\zeta \kappa (\kappa + 2 \eta^2 m^2) (3 \zeta G S_0^2 + 4 \pi^2\ell |J_0| )}}{m \sqrt[4]{3 G} \sqrt{\ell |\dot J_0| \big(3 \pi \zeta G S_0^2 - \pi^3 \ell \big(4 |J_0|+|\dot J_0| \tau  \big)\big)}} \nonumber \\
	&\times \left( \sqrt{3 \zeta G S_0^2 + 4 \pi^2 \ell |J_0|}-\sqrt{3 \zeta G S_0^2 + \pi^2 \ell \big(4 |J_0|-|\dot J_0| \tau \big)}\right),\quad \epsilon>0,
\end{align}
which is valid for $\epsilon > 0$. For small enough relaxation times we can expand these in Taylor series:
\begin{align}
	&\mathcal{L}_{\text{on-shell}}^{(+)}(\tau) \approx\frac{\pi ^{3/2}  \sqrt{-\epsilon\ell |\dot J_0|  }\, \sqrt[4]{\zeta  \kappa  \left(\kappa +2 \eta ^2 m^2\right)}}{m\sqrt[4]{3G} \left(3 \zeta  G S_0^2+4 \pi ^2 |J_0| \ell \right)^{3/4}}\,\tau+\mathcal{O}(\tau^2),\quad \epsilon<0,
	\\[5pt]
		&\mathcal{L}_{\text{on-shell}}^{(-)}(\tau) \approx\frac{\pi ^{3/2}  \sqrt{\epsilon\ell |\dot J_0|  }\, \sqrt[4]{\zeta  \kappa  \left(\kappa +2 \eta ^2 m^2\right)}}{m\sqrt[4]{3G} \left(3 \zeta  G S_0^2+4 \pi ^2 |J_0| \ell \right)^{3/4}}\,\tau+\mathcal{O}(\tau^2), \quad \epsilon>0.
\end{align}
The quantity in front of $\tau$ is called initial thermodynamic speed of the process. 
On the other hand, the unparameterized (direct) form of the lengths, for transitions between $J_0$ and $J_\tau$, are:
\begin{equation}
	\mathcal{L}_{\text{direct}}^{(\pm)} = \frac{\sqrt{\mp \epsilon} \sqrt[4]{\zeta \kappa (\kappa + 2 \eta^2 m^2)}}{m \sqrt[4]{3 G} \sqrt{\pi \ell}} \left| \sqrt[4]{3 \zeta G S_0^2 + 4 \pi^2 \ell J_\tau} - \sqrt[4]{3 \zeta G S_0^2 + 4 \pi^2 \ell J_0} \right|.
\end{equation}

Forcing the on-shell and the direct lengths to yield the same result, we can estimate the relaxation time for the isentropic process. For small times, one has:
\begin{align}
	&\tau\approx\frac{ \left(3 \zeta  G S_0^2+4 \pi ^2 |J_0| \ell \right)^{3/4} }{\pi ^{2} \ell \sqrt{  |\dot J_0|} }\left| \sqrt[4]{3 \zeta G S_0^2 + 4 \pi^2 \ell J_\tau} - \sqrt[4]{3 \zeta G S_0^2 + 4 \pi^2 \ell J_0} \right|.
\end{align}

Let us note that the thermodynamic geodesic equations in the $(S, J)$ state space do not admit solutions with constant angular momentum ($J = \text{const}$). As a consequence, within this geometric framework, the GMG warped black hole cannot maintain a steady state of rotation and must undergo a continuous change of its angular momentum. A complete characterization of this dynamical behavior requires a full non-perturbative numerical analysis of the coupled thermodynamic geodesic equations. 

Finally, if one wishes to investigate processes constrained to a constant mass-energy ($M = \text{const}$), the analysis must be transferred to the entropy representation of the GMG black hole thermodynamics.

\subsection{Geodesic equations in $(M,J)$ space and fluctuations}

In the entropy entropy ensemble one has entropy in its natural variables:
\begin{align}
&S(M,J)=\frac{ 6 m M \ell+\sqrt{\frac{6 \pi^2 \ell }{G \kappa }\left(\kappa +2 \eta ^2 m^2\right) \left(6\pi  G m^2 M^2 \ell -\zeta  \eta ^2 \kappa  J\right)}}{3 \zeta  \eta ^2 m},
\\[5pt]
&T(M,J)=\bigg(\frac{\partial S}{\partial M}\bigg)^{\!\!-1}_J=\frac{\zeta  \eta ^2 \sqrt{\kappa  \left(6 G m^2 M^2 \ell -\zeta  \eta ^2 \kappa  J\right)}}{2 \pi  \ell  \left(\sqrt{\kappa  \left(6 G m^2 M^2 \ell -\zeta  \eta ^2 \kappa  J\right)}+ m M \sqrt{6 G \ell  \left(\kappa +2 \eta ^2 m^2\right)}\right)}
\\[5pt]
&\Omega(M,J)=-T \frac{\partial S}{\partial J}\bigg|_M=\frac{\zeta  \eta ^2 \kappa  \sqrt{\kappa +2 \eta ^2 m^2}}{2 m \left( \sqrt{6G \kappa  \ell  \left(6 G m^2 M^2 \ell -\zeta  \eta ^2 \kappa  J\right)}+6 G m M \ell  \sqrt{\kappa +2 \eta ^2 m^2}\right)}.
\end{align}

Causality requires $S>0$ and $T>0$, which bounds the angular momentum:
\begin{equation}
J<\frac{6 G m^2 M^2 \ell }{\zeta  \eta ^2 \kappa }.
\end{equation}
Furthermore, since $C_J$ diverges at $J\to0$, we will consider only the $J>0$ sector. 
 
Since we are working in entropy representation, one has the Ruppeiner thermodynamic metric  defined by the Hessian of the entropy:
%^
\begin{align}
g^{R}=\epsilon	\mathcal{H}_S &= \!
	\begin{pmatrix}
		\left.\frac{\partial^2 S}{\partial M^2}\right|_J & \!\!\!\frac{\partial^2 S}{\partial M\partial J} \\[12pt]
		\frac{\partial^2 S}{\partial M\partial J} & \!\!\!\left.\frac{\partial^2 S}{\partial J^2}\right|_M
	\end{pmatrix} 
	\!=\! \left(
	\begin{array}{cc}
		\!\!\!-\frac{2\epsilon \sqrt{6} \pi  J m \sqrt{G \kappa  \ell ^3 \left(\kappa +2 \eta ^2 m^2\right)}}{\left(6 G m^2 M^2 \ell -\zeta  \eta ^2 \kappa  J\right)^{3/2}} & \!\!\!\frac{\sqrt{6} \epsilon\pi  m M \sqrt{G \kappa  \ell ^3 \left(\kappa +2 \eta ^2 m^2\right)}}{\left(6 G m^2 M^2 \ell -\zeta  \eta ^2 \kappa  J\right)^{3/2}} \\[12 pt]
		\!\!\!\frac{\sqrt{6}\epsilon \pi  m M \sqrt{G \kappa  \ell ^3 \left(\kappa +2 \eta ^2 m^2\right)}}{\left(6 G m^2 M^2 \ell -\zeta  \eta ^2 \kappa  J\right)^{3/2}} & \!\!\!-\frac{\pi \epsilon \zeta  \eta ^2 \sqrt{\kappa ^3 \ell  \left(\kappa +2 \eta ^2 m^2\right)}}{2 m \sqrt{6 G} \left(6 G m^2 M^2 \ell -\zeta  \eta ^2 \kappa  J\right)^{3/2}} \\
	\end{array}\!\!
	\right)\!\!.
\end{align}
Its curvature invariant does not constrain the ($M,J$) state space any further:
\begin{equation}
R^{R}=\frac{\sqrt{3} \zeta  \eta ^2 G \kappa  m}{\pi  \epsilon  \sqrt{2 G \kappa  \ell  \left(\kappa +2 \eta ^2 m^2\right) \left(6 G m^2 M^2 \ell -\zeta  \eta ^2 \kappa  J\right)}}.
\end{equation}

Under the Ruppeiner  metric $g^{(R)} = \epsilon \mathcal{H}_S$, the geodesic equations in ($M,J$) space are:
\begin{align}\label{eqSenera}
	&\ddot M + \Gamma^M_{MM} \dot M^2 + 2\Gamma^M_{M J} \dot M \dot J + \Gamma^M_{JJ} \dot J^2 = 0, \\[5pt]\label{eqJenera}
	&\ddot J + \Gamma^J_{JJ} \dot J^2 + 2\Gamma^J_{M J} \dot M \dot J + \Gamma^J_{MM} \dot M^2 = 0,
\end{align}
with the Christoffel symbols given by:
\begin{align}
	&\Gamma^{M}_{MM} =-\frac{6 G m^2 \ell  M}{6 G m^2 \ell  M^2- \zeta  \eta ^2 \kappa  J}, 
	\quad \Gamma^{M}_{MJ} = \frac{\zeta  \eta ^2 \kappa }{4(6 G m^2 \ell  M^2- \zeta  \eta ^2 \kappa  J)}, 
	\quad \Gamma^{M}_{JJ} = 0, \\[5pt]
	&\Gamma^{J}_{MM} = \frac{6 G m^2 \ell  J}{6 G m^2 \ell  M^2-\zeta  \eta ^2 \kappa  J} , 
	\quad \Gamma^{J}_{MJ} = \frac{6 G m^2 \ell  M}{\zeta  \eta ^2 \kappa  J-6 G m^2 \ell  M^2}
	, \quad \Gamma^{J}_{JJ} =\frac{3 \zeta  \eta ^2 \kappa }{4(6 G m^2 \ell  M^2-\zeta  \eta ^2 \kappa  J)} .
\end{align}

An nalytical solution exists for an isoenergetic (constant mass $M=M_0=\text{const}$) fluctuation. In this case, Eq.~\eqref{eqSenera} is satisfied identically, and Eq.~\eqref{eqJenera} reduces to:
\begin{equation}
	\ddot J(t) +\frac{3 \zeta  \eta ^2 \kappa }{24 G m^2 M_0^2 \ell -4 \zeta  \eta ^2 \kappa  J(t)}\dot J(t)^2 = 0, \quad J(0) = J_0, \quad \dot J(0) = \dot J_0,
\end{equation}
where $J_0$ and $\dot J_0$ denote the initial angular momentum and its rate of change. The solution is a fourth degree polynomial in  time:
\begin{align}
	J_{\pm}(t) &=J_0\pm|\dot J_0| t+\frac{3 \zeta  \eta ^2 \kappa  \dot J_0^2 }{8 \left(\zeta  \eta ^2 \kappa  J_0-6 G m^2 M_0^2 \ell \right)} t^2
	\\[5pt]
	&\pm\frac{\zeta ^2 \eta ^4 \kappa ^2 |\dot J_0|^3 }{16 \left(\zeta  \eta ^2 \kappa  J_0-6 G m^2 M_0^2 \ell \right)^2}t^3
	-\frac{\zeta ^3 \eta ^6 \kappa ^3 \dot J_0^4 }{256 \left(6 G m^2 M_0^2 \ell -\zeta  \eta ^2 \kappa  J_0\right)^3} t^4,
\end{align}
where, depending on the sign of the initial rate $\dot J_0$,we have two solutions $J_{\pm}$.

Inserting these solutions into the thermodynamic length functional, we find:
\begin{align}
	\mathcal{L}_{\text{on-shell}}^{(+)}(\tau) &= 4 \sqrt{-\pi  \epsilon }\frac{ \sqrt[4]{2 \ell  \left(\kappa +2 \eta ^2 m^2\right) \left(6 G m^2 M_0^2 \ell -\zeta  \eta ^2 \kappa  J_0\right)}}{\eta  \sqrt[4]{3 G \kappa } \,\sqrt{\zeta m |\dot J_0| \big(24 G m^2 M_0^2 \ell -\zeta  \eta ^2 \kappa  \big(4 J_0+|\dot J_0| \tau\big)\big)}} \nonumber \\[5pt]
	&\times \left(2 \sqrt{6 G m^2 M_0^2 \ell -\zeta  \eta ^2 \kappa  J_0}-\sqrt{24 G m^2 M_0^2 \ell -\zeta  \eta ^2 \kappa  \big(4 J_0+|\dot J_0| \tau \big)}\right),
\end{align}
valid for $\epsilon < 0$ and:
\begin{align}
	\mathcal{L}_{\text{on-shell}}^{(-)}(\tau) &= 4 \sqrt{\pi  \epsilon }\frac{ \sqrt[4]{2 \ell  \left(\kappa +2 \eta ^2 m^2\right) \left(6 G m^2 M_0^2 \ell -\zeta  \eta ^2 \kappa  J_0\right)}}{\eta  \sqrt[4]{3 G \kappa } \,\sqrt{\zeta m |\dot J_0|  \left(24 G m^2 M_0^2 \ell -\zeta  \eta ^2 \kappa  \big(4 J_0-|\dot J_0| \tau\big)\right)}} \nonumber \\[5pt]
	&\times \left(\sqrt{24 G m^2 M_0^2 \ell -\zeta  \eta ^2 \kappa  \big(4 J_0-|\dot J_0| \tau \big)}-2 \sqrt{6 G m^2 M_0^2 \ell -\zeta  \eta ^2 \kappa  J_0}\right),
\end{align}
which is valid for $\epsilon > 0$. For small enough relaxation times we can expand these in Taylor series:
\begin{align}
	&\mathcal{L}_{\text{on-shell}}^{(+)}(\tau) \approx\eta \, \sqrt{-\frac{\pi  \zeta \epsilon |\dot J_0| }{m}} \frac{   \sqrt[4]{\kappa ^3 \ell  \left(\kappa +2 \eta ^2 m^2\right)}}{\sqrt[4]{24 G} \left(6 G m^2 M_0^2 \ell - \zeta  \eta ^2 \kappa  J_0\right)^{3/4}} \,\tau +\mathcal{O}(\tau^2),\quad \epsilon<0,
	\\[5pt]
	&\mathcal{L}_{\text{on-shell}}^{(-)}(\tau) \approx\eta \, \sqrt{\frac{\pi  \zeta \epsilon |\dot J_0| }{m}} \frac{   \sqrt[4]{\kappa ^3 \ell  \left(\kappa +2 \eta ^2 m^2\right)}}{\sqrt[4]{24 G} \left(6 G m^2 M_0^2 \ell - \zeta  \eta ^2 \kappa  J_0\right)^{3/4}} \,\tau +\mathcal{O}(\tau^2), \quad \epsilon>0.
\end{align}
The quantity in front of $\tau$ is called initial thermodynamic speed of the process. 
On the other hand, the unparameterized (direct) form of the lengths, for transitions between $J_0$ and $J_\tau$, are:
\begin{equation}
	\mathcal{L}_{\text{direct}}^{(\pm)} = \frac{2 \sqrt{\pi  \epsilon } \sqrt[4]{2 \ell  \left(\kappa +2 \eta ^2 m^2\right)}}{\eta  \sqrt{\zeta  m} \sqrt[4]{3 G \kappa }}\left| \sqrt[4]{6 G m^2 M^2 \ell -\zeta  \eta ^2 \kappa  J_0}-\sqrt[4]{6 G m^2 M^2 \ell -\zeta  \eta ^2 \kappa  J_{\tau }} \right|.
\end{equation}

Forcing the on-shell and the direct lengths to yield the same result, we can estimate the relaxation time for the isentropic process. For small times, one has:
\begin{align}
	&\tau\approx\frac{4 \left(6 G m^2 M_0^2 \ell -\zeta  \eta ^2 \kappa  J_0\right)^{3/4}}{\zeta  \eta ^2 \kappa  \sqrt{|\dot J_0|}}\left| \sqrt[4]{6 G m^2 M^2 \ell -\zeta  \eta ^2 \kappa  J_0}-\sqrt[4]{6 G m^2 M^2 \ell -\zeta  \eta ^2 \kappa  J_{\tau }} \right|.
\end{align}

Let us note that the thermodynamic geodesic equations in the $(M, J)$ state space also do not admit solutions with constant angular momentum ($J = \text{const}$).

\section{Conclusion}\label{secConcl}

In this work, we have systematically investigated the classical thermodynamic stability of the warped black hole solution in three-dimensional General Massive Gravity (GMG). Our analysis demonstrates that the GMG warped black hole is thermodynamically unstable, indicating that it cannot exist in a genuine equilibrium state. Consequently, the system is expected to undergo continuous radiation-driven evolution. This behavior persists in the appropriate limit to the warped AdS$_3$ black hole of Topological Massive Gravity, where the same instability is recovered.

At the level of local thermodynamic stability, the analysis of the admissible heat capacities revealed  nontrivial Davies phase-transition curves. These occur in the asymptotic regimes $\zeta\to 0$ and $\zeta\to \infty$, the static limit $J\to0$, and also the extremal boundaries $J\to J_{\mathrm{extr}}^{(\pm)}$, defined in Eq.~\eqref{eqExtrJGMG}. The presence of these singular loci signals qualitative changes in the thermodynamic behavior of the system.

To further investigate the nonequilibrium properties of the GMG warped black hole, we employed the framework of stochastic thermodynamic geometry based on Ruppeiner's fluctuation theory, using either a suitably rescaled Weinhold metric, $g^{(W)}=\epsilon\mathcal{H}_M$, or a rescaled Ruppeiner metric, $g^{(R)}=\epsilon\mathcal{H}_S$. In the Weinhold formulation, we focused on isentropic fluctuations ($S=\mathrm{const}$) in the $(S,J)$ thermodynamic state space and derived exact analytical expressions for the on-shell angular momentum trajectories and the associated thermodynamic lengths. These quantities characterize the relative probabilities of relaxation processes along specific paths connecting macrostates. Furthermore, a Taylor expansion of the thermodynamic lengths enabled us to estimate the characteristic relaxation times of the corresponding isentropic processes.

An analogous analysis was performed in the $(M,J)$ state space using the Ruppeiner metric as the underlying thermodynamic geometry. In this case, we considered isoenergetic processes ($M=\mathrm{const}$) and obtained qualitatively similar results, including exact on-shell trajectories, thermodynamic lengths, and estimates of the corresponding relaxation times.

A notable outcome of our analysis is that the geodesic equations, governing the thermodynamic fluctuations in both $(S,J)$ and $(M,J)$ state space, do not admit solutions with constant angular momentum, $J=\mathrm{const}$. Within the geometric framework considered here, this suggests that the GMG warped black hole cannot sustain a stationary rotational state and must instead undergo a continuous evolution of its angular momentum as it relaxes between thermodynamic configurations.

Several directions remain open for future investigation. A complete characterization of fluctuations along arbitrary paths in the thermodynamic state space will require a fully non-perturbative numerical treatment of the coupled thermodynamic geodesic equations. In addition, it would be interesting to extend the present study to non-Hessian formulations of thermodynamic geometry, including alternative geometric approaches, \cite{Quevedo:2007ws, Quevedo:2007mj, Quevedo:2017tgz, pineda2019physical}, that may provide complementary insights into the stability, fluctuation structure, and relaxation dynamics of the GMG warped black hole solution.

\section*{Acknowledgments}
The author would like to express his gratitude to R. C. Rashkov, H. Dimov, M. Radomirov, V. Avramov and  S. Yazadjiev for their invaluable
comments and discussions. T. V. is fully financed by the European Union-NextGeneration EU, through the
National Recovery and Resilience Plan of the Republic of Bulgaria, project BG-RRP-2.004-0008-C01.

\bibliographystyle{utphys}
%\bibliography{References}

\providecommand{\href}[2]{#2}\begingroup\raggedright\endgroup

\end{document}